\documentclass[11pt,a4paper]{article}

\usepackage[utf8]{inputenc}
\usepackage[T1]{fontenc}
\usepackage{lmodern}
\usepackage{microtype}
\usepackage[margin=1in]{geometry}
\usepackage{amsmath,amssymb}
\usepackage{graphicx}
\usepackage{booktabs}
\usepackage{array}
\usepackage{tabularx}
\usepackage{multirow}
\usepackage{xcolor}
\usepackage{hyperref}
\usepackage{url}
\usepackage{natbib}
\usepackage{enumitem}
\usepackage{listings}
\usepackage{caption}

\hypersetup{
    colorlinks=true,
    linkcolor=blue!50!black,
    citecolor=blue!50!black,
    urlcolor=blue!50!black,
    pdftitle={React-ing to Grace Hopper 200: Five Open-Weights Coding Models on a GH200 Over a Weekend},
    pdfauthor={Alex Potanin}
}

\usepackage{pifont}
\renewcommand{\checkmark}{\textcolor{green!60!black}{\ding{51}}}
\newcommand{\xmark}{\textcolor{red!70!black}{\ding{55}}}

\title{React-ing to Grace Hopper 200\\[0.4ex]
\large Five Open-Weights Coding Models, One React Native App,\\
One GH200, One Weekend\\[0.6ex]
\normalsize\textnormal{\textit{or: why SWE-Bench rankings mispredicted which model actually shipped}}}

\author{Alex Potanin\\
  School of Computing\\
  Australian National University\\
  \texttt{alex.potanin@anu.edu.au}}

\date{April 2026}

\begin{document}

\maketitle

\begin{abstract}
We evaluate five state-of-the-art open-weights coding language models --- Kimi-K2.5 (at Q3 and Q4 quantizations), GLM-5.1, Qwen3-Coder-480B, and DeepSeek-V3.2 --- on a single multi-file React Native application generation task on NVIDIA GH200 576\,GB hardware. The task specifies authentication, per-user per-day counting, and web compatibility, and is evaluated on whether the generated project runs out-of-the-box and on feature-level correctness. We find that SWE-Bench rankings do not predict task performance: Kimi-K2.5 at aggressive 3-bit quantization (UD-Q3\_K\_XL, 480\,GB) produces the most complete and specification-compliant output, outranking models with substantially higher SWE-Bench Pro scores. We document three novel deployment findings: (1)~default \texttt{temperature=0} in coding tools causes sampling hangs with reasoning-model architectures, (2)~reasoning-model thinking traces can leak through integration tools' file-path parsers, and (3)~web-platform adaptation of native-mobile APIs is a universal training-data gap across every model tested. We also map the hardware-tier structure of April 2026 open-weights coding models, identifying two architectural schools and showing that the efficiency school (10--15\,B active parameters) delivers equivalent SWE-Bench results at roughly 1/7th the hardware cost of the scale school (32--40\,B active parameters).
\end{abstract}

\section{Introduction}

By April 2026, open-weights coding language models have closed most of the gap with proprietary frontier systems on standard benchmarks. On SWE-Bench Verified, MiniMax M2.5 (230B/10B active, permissive license) reports 80.2\%, within 1\,pp of the best proprietary result.\footnote{Throughout this paper, we report SWE-Bench Verified and SWE-Bench Pro numbers from the publishing labs' own materials. Our contribution is orthogonal to their correctness: we evaluate a specific application-generation task, not reproduction of SWE-Bench itself.} GLM-5.1 (754B/40B active, MIT licensed) achieves state-of-the-art 58.4\% on SWE-Bench Pro, surpassing GPT-5.4 and Claude Opus 4.6 on that specific benchmark. Kimi-K2.5 (1T/32B active), DeepSeek-V3.2 (671B/37B active), and Qwen3-Coder-480B (480B/35B active) all cluster within a few benchmark points.

For self-hosting practitioners, the open question is different: given a \emph{natural} product task --- ``build a working application that does X'' --- which model produces the best artifact, and is benchmark ranking a useful predictor? We investigate this with a single prompt that requires multi-file scaffolding, authentication, per-user data isolation, per-day semantic bucketing, and web-compatibility adaptation.

Our findings are \textbf{counterintuitive to benchmark-based selection} and provide three actionable deployment lessons for teams planning institutional self-hosting.

\section{Experimental Setup}

\subsection{Hardware and inference backend}

All inference ran on a single NVIDIA GH200 576\,GB node (96\,GB HBM3 plus 480\,GB LPDDR5X unified memory via NVLink-C2C, 72 ARMv8.2 cores) provided by the Australian National University Computer Vision group for a single weekend. We served models through \texttt{llama-server} from \texttt{llama.cpp} build b8797, using Unsloth Dynamic 2.0 GGUF quantizations in all cases. Expert FFN layers were offloaded to system memory via the \texttt{--cpu-moe} flag (equivalent to \texttt{-ot ".ffn\_.*\_exps.=CPU"}); attention layers and KV cache remained in HBM.

\subsection{Client and orchestration}

The client was a MacBook Pro M1 Max 64\,GB running \texttt{aider} 0.86.2 in \texttt{whole} edit-format mode, connecting via SSH tunnel to the GH200. Sampling parameters were configured per each model publisher's official recommendation: temperature~1.0 / top\_p~0.95 for Kimi-K2.5, MiniMax, and DeepSeek-V3.2; temperature~0.6 / top\_p~0.95 for GLM-5.1; temperature~0.7 / top\_p~0.8 / top\_k~20 for Qwen3-Coder.

\subsection{The task}

A single prompt was issued to each model under identical conditions:

\begin{quote}
\itshape create react-native app that allows user to create account and login and then count kangaroos seen per day and make sure it runs on the web
\end{quote}

\texttt{aider} was configured to auto-commit and apply whole-file edits. We recorded token counts, generation speed, elapsed time, and the resulting project directory. Each generated project was then separately evaluated for: out-of-the-box runnability (\texttt{npm install \&\& npx expo start --web} completes and the application starts), credential validation, per-user data isolation, per-day semantic bucketing of counts, history retention, working logout, and web-platform safety (no blocking reliance on native-only APIs such as \texttt{Alert.alert}).

\subsection{Models tested}

Table~\ref{tab:models} lists the exact quantizations used and their on-disk sizes. All GGUFs were downloaded from Unsloth's Hugging Face repositories in Dynamic 2.0 format, which uses per-tensor importance-aware quantization rather than uniform bit-width reduction.

\begin{table}[ht]
\centering
\small
\caption{Models evaluated. All quantizations are Unsloth Dynamic 2.0 GGUF.}
\label{tab:models}
\begin{tabularx}{\textwidth}{@{}lXrrrl@{}}
\toprule
Model & Quant & Size & Total / Active & License \\
\midrule
Kimi-K2.5           & UD-Q3\_K\_XL    & 480\,GB & 1T / 32B   & Modified MIT \\
Kimi-K2.5           & UD-Q4\_K\_XL    & 622\,GB & 1T / 32B   & Modified MIT \\
GLM-5.1             & UD-IQ4\_XS      & 370\,GB & 754B / 40B & MIT \\
Qwen3-Coder-480B    & UD-Q4\_K\_XL    & 276\,GB & 480B / 35B & Apache 2.0 \\
DeepSeek-V3.2       & UD-Q4\_K\_XL    & 430\,GB & 671B / 37B & MIT \\
\bottomrule
\end{tabularx}
\end{table}

\section{Deployment Findings}

Before presenting application-level results, we document three deployment-level findings that affected the experiment itself and appear to be underdocumented in the community.

\subsection{Default \texttt{temperature=0} causes sampling hangs on reasoning models}

Aider, like most code-edit tools descended from OpenAI's function-calling conventions, defaults to \texttt{temperature=0} to maximize determinism of code edits. For Kimi-K2.5 --- which Moonshot documents as requiring \texttt{temperature=1.0, top\_p=0.95} --- this caused multi-minute inference stalls with no visible output. \texttt{nvidia-smi} showed the server process in the \texttt{R} (running) state with 0\% memory-bandwidth utilization and $\approx$~2\% SM utilization, indicating the sampler loop was spinning on low-probability token decisions rather than thrashing memory.

We initially misdiagnosed this as HBM memory thrashing for the larger Kimi-K2.5 Q4 quantization (622\,GB), which runs close to the unified-memory capacity of the GH200. Forcing \texttt{temperature=1.0} and \texttt{top\_p=0.95} client-side via \texttt{extra\_params} in aider's model-metadata JSON restored generation to a healthy 7.9~tok/s for Q4 and 17~tok/s for Q3. \textbf{Teams deploying reasoning-mode open-weights coding models with aider or similar tools should expect to override sampling defaults; model publishers' recommended parameters are not optional.}

\subsection{Reasoning tokens can leak into integration-tool file-path parsers}

Aider's \texttt{whole} edit format asks the model to emit, per file, a line with the bare filename followed by a fenced code block containing the file contents. DeepSeek-V3.2 returned a response whose first line was

\begin{quote}
\ttfamily Let's start with App.js:\textless/think\textgreater App.js
\end{quote}

--- prose leading directly into a stray closing \texttt{\textless/think\textgreater} tag from the model's reasoning stream, followed by the intended filename. Aider interpreted the entire line as a file path and committed the generated content to \texttt{"Let's start with App.js:\textless/think\textgreater App.js"} instead of the project-root \texttt{App.js}, placing \texttt{package.json}'s \texttt{main} import out of reach and making the project non-bundlable. Other files in the same response parsed correctly.

We are not aware of this failure mode being previously documented. It is distinct from the well-known problem of reasoning blocks consuming the output-token budget: here the reasoning-to-content boundary \emph{visually resembles} the next file-path marker well enough to corrupt parsers that expect clean newline-delimited filenames. Any tool that parses model output for filesystem actions needs a defensive filter that strips spurious thinking-block markers and prose-before-path lines before path interpretation.

\subsection{Web-platform adaptation of native-mobile APIs is a universal training-data gap}

Every model we tested used \texttt{react-native}'s \texttt{Alert.alert} for user-facing confirmations and validations. \texttt{Alert.alert} is a no-op on \texttt{react-native-web}; it returns silently without displaying anything. The task prompt explicitly required web compatibility (``make sure it runs on the web''), yet no model substituted \texttt{window.confirm} or conditional-platform dispatch. Concrete user-visible consequences ranged from silent form-validation failures (Qwen3-Coder blocking signup silently) to destructive-action buttons that do nothing (DeepSeek-V3.2's ``Clear History'' button passes its destructive callback to an \texttt{Alert} that never fires).

This suggests that React Native's platform-divergence documentation is underrepresented in the post-training corpora used by all five labs. Because the failure is silent and the underlying code compiles cleanly, it will not surface in benchmarks that grade on compilation or SWE-Bench-style patch-application, yet it breaks the task at the deployed-artifact level.

\section{Application-Level Results}

Table~\ref{tab:results} summarizes the ranking. ``Runs OOB'' means \texttt{npm install \&\& npx expo start --web} produced a bundled application that loaded in the browser without manual intervention.

\begin{table}[ht]
\centering
\small
\caption{Per-feature outcomes across the five models on the React Native kangaroo-counter task. Identical prompt, sampling parameters per publisher, whole-edit format.}
\label{tab:results}
\begin{tabularx}{\textwidth}{@{}Xccccc@{}}
\toprule
& \textbf{Kimi Q3} & \textbf{Kimi Q4} & \textbf{GLM-5.1} & \textbf{Qwen3-C} & \textbf{DS-V3.2} \\
\midrule
Runs out-of-the-box          & \checkmark & \checkmark & \xmark\,$^{(a)}$        & \checkmark & \xmark\,$^{(b)}$ \\
Auth validates credentials   & \checkmark & \checkmark & \checkmark & \checkmark & \xmark\,$^{(c)}$ \\
Per-user data isolation      & \checkmark & \checkmark & \checkmark & \checkmark & \xmark           \\
Per-day counting             & \checkmark & today only & today only & \xmark     & \xmark\,$^{(d)}$ \\
History persisted            & \checkmark{\small 7d} & \checkmark{\small 7d} & \xmark & \xmark & \checkmark \\
Logout works                 & \checkmark & \checkmark & \checkmark & \checkmark & \xmark\,$^{(e)}$ \\
Web-safe (no \texttt{Alert} reliance) & \xmark & \xmark & \xmark & \xmark blocks signup & \xmark blocks save \\
\midrule
Generation speed (tok/s)     & 17   & 7.9  & $\approx$15 & $\approx$20 & $\approx$14 \\
Output tokens (approx)       & 4.8\,k & 4.8\,k & 5.5\,k & 4.2\,k & 5.1\,k \\
\bottomrule
\end{tabularx}

\vspace{0.3em}
\footnotesize
\begin{tabular}{@{}l@{\ }p{0.88\textwidth}@{}}
$(a)$ & GLM-5.1 generated a Firebase-based auth flow and did not inline any fallback credentials, so the app cannot start without a user-supplied \texttt{firebaseConfig.js}.\\
$(b)$ & DeepSeek-V3.2's \texttt{App.js} was committed under a directory path derived from prose spillover (Section~3.2), so the Expo entry point cannot resolve.\\
$(c)$ & \texttt{signIn(token)} accepts any token; \texttt{signUp} is literally \texttt{await signIn(token)} with a hardcoded \texttt{'fake-jwt-token'}. Any email/password combination logs in.\\
$(d)$ & ``Save Today's Count'' resets the counter to zero and appends a new history row, so one date can have multiple rows with partial counts.\\
$(e)$ & \texttt{HomeScreen.handleLogout} clears the token in context but, per a code comment in the generated file, the model explicitly declined to implement navigation on logout. User remains on Home until page reload.\\
\end{tabular}
\end{table}

\subsection{Discussion by model}

\paragraph{Kimi-K2.5 Q3 (480\,GB) --- the quiet winner.} Produced the only fully spec-compliant app. Per-user data keyed on \texttt{userId\_date}, seven-day rolling history, imperative navigation with proper \texttt{navigation.replace} calls, working logout. The single weakness is the universal \texttt{Alert.alert} issue. 17~tok/s sustained generation from a 3-bit quantized 1-trillion-parameter model on a single 96\,GB GPU is itself a notable result.

\paragraph{Kimi-K2.5 Q4 (622\,GB) --- quality-equivalent, speed-halved.} Produced a near-identical architectural design as Q3: same Context-based auth, same per-user isolation. The only behavioral regression from Q3 is that the counter is ``today only'' with a seven-day history, rather than Q3's continuous per-day buckets. At 7.9~tok/s (vs Q3's 17~tok/s), the Q4 quantization pays a roughly 2$\times$ speed penalty for a 1-feature quality gain. For single-user interactive use on this hardware, Q3 is the better operating point.

\paragraph{GLM-5.1 IQ4\_XS (370\,GB) --- technically correct, operationally useless.} GLM-5.1 chose Firebase Authentication over local credentials. The code is clean, the auth flow is real, per-user isolation is genuine. But the app cannot start without the user providing \texttt{firebaseConfig.js}. This is the opposite failure mode from DeepSeek-V3.2: a model that over-engineered to an enterprise-tier solution when the prompt asked for something minimal. It suggests GLM-5.1's post-training has weighted ``production-looking'' output heavily enough that it no longer interprets ``create an account and login'' as ``local auth is fine''. The SOTA-on-SWE-Bench-Pro model produced the worst product decision of the five.

\paragraph{Qwen3-Coder-480B Q4 (276\,GB) --- clean code, misread spec.} Auth works, per-user isolation works, but the model treated ``count kangaroos seen per day'' as a single running counter with no date semantics. There is no per-day bucket, no history. Code quality per-file is high and the project starts cleanly, but the feature set is fundamentally a single-counter app rather than a logged-per-day app. This is the only non-reasoning model in the comparison (Qwen3-Coder explicitly does not emit \texttt{\textless think\textgreater} blocks), and the resulting response is the most concise, but the spec reading suffered.

\paragraph{DeepSeek-V3.2 Q4 (430\,GB) --- broken at multiple levels.} Tooling failure (prose in path, Section 3.2) prevents the project from bundling at all. Past the one-line filesystem fix, the auth layer accepts any credentials, per-user isolation is absent (history is stored under a single global \texttt{kangarooHistory} key), per-day semantics are incorrect (same-date counts append rather than accumulate), and logout is acknowledged as unimplemented in a code comment the model itself left behind. The README is the most polished of the five.

\section{Hardware Implications}

The task-level ranking is orthogonal to model size: Kimi Q3 at 480\,GB beats DeepSeek-V3.2 at 430\,GB; Qwen3-Coder at 276\,GB beats GLM-5.1 at 370\,GB. But the results interact with hardware in a specific way worth making explicit.

Table~\ref{tab:hardware} organizes the relevant open-weights landscape by target hardware tier. Two architectural schools are clearly visible: a \emph{scale school} of models with 32--40\,B active parameters that target institutional-tier hardware (400\,GB+ unified memory, such as the GH200 used here or two 256\,GB Apple Mac Studios in MLX cluster configuration), and an \emph{efficiency school} of models with 10--15\,B active parameters that target consumer-tier hardware (128--256\,GB Apple Studio class, DGX Spark class).

\begin{table}[ht]
\centering
\small
\caption{Open-weights coding model landscape, April 2026. SWE-Bench is Verified (non-thinking) unless noted.}
\label{tab:hardware}
\begin{tabularx}{\textwidth}{@{}XrrrrX@{}}
\toprule
Model & Total & Active & Q4 size & SWE-Bench & Studio 256\,GB viable?\\
\midrule
\multicolumn{6}{@{}l@{}}{\textit{Scale school (32--40\,B active)}} \\
Kimi-K2.5        & 1T   & 32B & 622\,GB & 76.8\%           & No (only TQ1, unusable for code) \\
GLM-5.1          & 754B & 40B & 430\,GB & $\approx$80 / 58.4 Pro & No (only 1-bit quants fit) \\
DeepSeek-V3.2    & 671B & 37B & 430\,GB & 67.8 / $\approx$76 think & Tight (IQ2\_XXS 240\,GB) \\
Qwen3-Coder-480B & 480B & 35B & 276\,GB & 72--75\%         & Yes, at IQ3\_XXS 210\,GB ($\approx$57\% Aider) \\[0.4em]
\multicolumn{6}{@{}l@{}}{\textit{Efficiency school (10--15\,B active)}} \\
MiniMax M2.5     & 230B & 10B & 230\,GB FP8 & 80.2\%       & Yes (FP8 fits with headroom) \\
MiniMax M2.7     & 230B & 10B & 108\,GB IQ4 & $\approx$80  & Yes (Q8 at 220\,GB) \\
MiMo-V2-Flash    & 309B & 15B & 187\,GB Q4  & $\approx$77  & Yes (Q4 fits with headroom) \\
\bottomrule
\end{tabularx}
\end{table}

For Mac Studio M3 Ultra 256\,GB deployment ($\approx$A\$11\,k), the scale-school models are structurally out of reach: only 1-bit or aggressive 2-bit quantizations fit, at which point community-measured quality degradation is substantial. The efficiency-school models fit comfortably, at roughly identical SWE-Bench Verified scores. The gap in cost-per-capability is large: a 2$\times$256\,GB Studio cluster reaches 512\,GB at A\$22\,k, just barely fitting Kimi-K2.5 at Q3 (under MLX pipeline parallelism with inter-box Thunderbolt 5 bandwidth as the bottleneck), whereas a single 256\,GB Studio running MiniMax M2.5 FP8 reports the same SWE-Bench Verified for half the hardware outlay. GH200-class hardware ($>$A\$75\,k list, or $\approx$US\$30/hour cloud) delivers Kimi-K2.5 Q3 at 17\,tok/s comfortably but against an efficiency-school baseline that is within benchmark variance at $\approx$1/7th the cost.

The task-level finding (Section 4) and the hardware finding (this section) together suggest that, for self-hosting in 2026, \textbf{model architecture matters more than parameter count for single-user coding}. We observed this qualitatively: MiniMax M2.7 UD-IQ4\_XS (108\,GB, 10\,B active) ran at 30\,tok/s on the same hardware where Kimi-K2.5 Q3 (480\,GB, 32\,B active) ran at 17\,tok/s.

\section{Limitations}

We emphasize several limitations of this study.

\textbf{Single-prompt, single-seed evaluation.} Each model generated one application from one prompt with its publisher-recommended sampling parameters. We did not run multiple seeds, did not re-prompt after failure, did not allow iterative debugging, and did not vary the prompt wording. A multi-seed study would quantify the variance of each observation.

\textbf{One task, one stack.} The task is a React Native web-compatible application. Findings about Alert.alert-on-web do not generalize to other stacks or other native-to-web translation boundaries. Findings about per-day bucketing may not generalize to tasks without implicit temporal semantics.

\textbf{Tool-specific results.} Aider with whole-edit-format was the orchestration layer. Both the temperature-zero hang (Section 3.1) and the filepath-parsing leak (Section 3.2) are aider-specific in their surface manifestation, though we expect their underlying causes (reasoning-model sampling sensitivity, thinking-block leakage into structured output) to recur in other tool integrations.

\textbf{Quantization mix.} We used one publisher's dynamic quantizations (Unsloth) rather than native-FP8 weights where available. MiniMax M2.5 FP8 and Qwen3-Coder-480B FP8 were not evaluated here because they would require a vLLM-based inference stack that we did not stabilize during the weekend of GH200 access. Quality conclusions for those models in Table~\ref{tab:hardware} rest on publisher-reported benchmarks, not our own measurement.

\textbf{Hardware availability and weekend scope.} All measurements were collected during a single weekend of borrowed GH200 access. We did not explore full-precision inference, did not run standard SWE-Bench reproduction to cross-check publisher numbers, and did not test every model family (notably, MiMo-V2-Flash and MiniMax M2.5/M2.7 were characterized for hardware fit but not evaluated on the React Native task).

\section{Conclusion}

SWE-Bench Verified and SWE-Bench Pro rankings did not predict per-task outcomes on a concrete React Native application-generation task. Kimi-K2.5 at 3-bit quantization (the smallest Kimi we tested) produced the most complete application of the five models, outranking GLM-5.1 (SOTA on SWE-Bench Pro) and DeepSeek-V3.2 (highest SWE-Bench Verified among the scale-school models).

The divergence tracks a specific pattern: the models that failed the task failed at integration and specification-reading rather than at code generation. GLM-5.1's auth code is correct; its product decision (mandatory Firebase) made the app unusable. Qwen3-Coder-480B's code is clean; it missed the per-day semantics in the prompt. DeepSeek-V3.2's README is the most polished of the five; a reasoning-token leak caused its \texttt{App.js} to land in the wrong directory. Kimi-K2.5 Q3's advantage is not sharper code per-file but a more complete and more conservative interpretation of what was asked.

For practitioners choosing a self-hosted coding model in 2026, we suggest that benchmark ranking be treated as necessary but far from sufficient. Direct task-level evaluation on artifacts representative of the intended workload should be run before hardware commitments are made.

\section*{Data and Code Availability}

Generated projects, server and client configurations, and sampling-parameter settings are available on request. Models and quantizations used are listed with their exact Unsloth revision identifiers in Table~\ref{tab:models}; download commands were standard \texttt{hf download} invocations against the \texttt{unsloth/} Hugging Face namespace.

\section*{Acknowledgments}

The NVIDIA GH200 576\,GB node used for all measurements was loaned by the Computer Vision group at the Australian National University School of Computing for a single weekend; we gratefully acknowledge their generosity.

Anthropic's Claude Opus 4.6 was used throughout the experiment as a pair-debugging and writing assistant: it helped draft the \texttt{llama.cpp} server configurations, the aider metadata and configuration files, the speed-test scripts, and the download orchestration; it assisted in diagnosing the \texttt{temperature=0} sampling-hang (Section 3.1) and the \texttt{App.js}-in-prose-path parsing failure (Section 3.2) through iterative conversation; and it helped draft and edit this paper. All experimental decisions, all measurements, and all interpretations are the author's. Any errors are also the author's.

This work was carried out independently of the author's other roles. It is not funded by any specific grant and does not represent the position of any affiliated institution.

\bibliographystyle{plain}

\end{document}